# Flow electrification of corona-charged polyethylene terephthalate film

Rui Kou,[1] Ying Zhong,[2,3] Yu Qiao[1,3,*]

[1] *Department of Structural Engineering, University of California – San Diego, La Jolla, CA 92093-0085, United States*

[2] *Department of Mechanical Engineering, University of South Florida, Tampa, Florida 33620, United States*

[3] *Program of Materials Science and Engineering, University of California – San Diego, La Jolla, CA 92093, United States*

[*] *Corresponding author. Email:* yqiao@ucsd.edu

**Abstract:** Corona charging a free-standing polymer film can produce a quasi-permanent potential difference across the film thickness, while the absolute amplitude of surface voltage may be highly sensitive to the free charges. To precisely control the voltage distribution, we investigated the flow electrification technology, by exposing corona-charged polyethylene terephthalate films to a variety of sodium salt solutions. The surface voltage and the free charge density were adjusted by the salt concentration, the anion size, and the flow rate. The dipolar component of electric potential remained unchanged. This result has significant scientific interest and technological importance to surface treatment, filtration, energy harvesting, bio-actuation and bio-sensing, among others.

*Keywords*: Corona charging; Flow electrification; Electrostatic; Polymer; Surface

## Nomenclature

| | |
|---|---|
| $e$ | Electron charge |
| $k_B$ | Boltzmann constant |
| $N_A$ | Avogadro's number |
| $m$ | Molarity of electrolyte (in mM) |
| PET | polyethylene terephthalate |
| $T$ | Temperature |
| $V_d$ | Dipolar component of surface voltage |
| $V_f$ | Surface voltage of free-standing film |
| $V_{f,N}$ | $V_f$ measured from the negative side |
| $V_{f,P}$ | $V_f$ measured from the positive side |
| $V_{Grid}$ | Voltage applied on the steel-wire grid |
| $V_{Needle}$ | Voltage applied on the needle electrode |
| $v_0$ | Liquid flow rate |
| $z_s$ | Distance of shear plane to the film surface (in nm) |
| $\Delta V_f$ | Voltage difference between $V_{f,P}$ and $V_{f,N}$ |
| $\varepsilon_0$ | Permittivity of vacuum |
| $\varepsilon_r$ | Relative permittivity |
| $\rho$ | Spatial charge density (in C/m$^3$) |
| $\sigma_d$ | Dipolar component of surface charge density (in C/m$^2$) |
| $\sigma_f$ | Density of free surface charges (in C/m$^2$) |
| $\sigma_{f,b}$ | Balanced density of free surface charges (in C/m$^2$) |
| $\sigma_N$ | Surface charge density on the negative side (in C/m$^2$) |
| $\sigma_P$ | Surface charge density on the positive side (in C/m$^2$) |
| $\psi_0$ | Electric potential at the solid surface |

## 1. Introduction

Polymer films can be quasi-permanently equipped with electric charges [1–3]. Charged polymer films have a wide range of applications in xerography and printing, filtration and sanitation, electrostatic actuation, micro-electromechanical systems, bio-sensing and bio-actuation, electronics, advanced acoustic devices, energy harvesting, to name a few [4–13]. Corona charging, as a fast, economical, and highly effective electrification technique, has been extensively studied [14–16]. It has been known that a corona-charged free-standing polymer film tends to have a stable potential difference across its thickness, which will be referred to as the dipolar component of electric potential ($V_d$) in the discussion below [17]. Yet, a relatively small amount of free surface charges can cause a large variation in the absolute amplitude of surface voltage ($V_f$), even when the density of free charges ($\sigma_f$) is less than the injected dipolar component of charge density ($\sigma_d$) by more than three orders of magnitude [1,17,18]. To achieve a stable overall electrification effect, not only $\sigma_d$ but also $\sigma_f$ need to be precisely controlled.

Upon contact with a liquid phase, e.g., an electrolyte solution, surface groups of a polymer may be ionized or dissociated, rendering its surface negatively charged. The surface charge would interrupt the diffuse layer, wherein the thermal disarray of ions balances the electric field [19][20]. When the liquid flows, the surface liquid layer within the shear plane would remain while the bulk liquid phase is removed [21–24], so that $V_f$ changes accordingly. In the current study, the liquid flow process was investigated systematically by using a set of sodium salt solutions, and its influence on $V_f$ of corona-charged polymer films was analyzed in detail.

## 2. Experimental procedure

A triode corona charging system was set up, as shown in Figure 1(a,b). The system consisted of a 1.5-mm-diameter 19-mm-long tungsten needle, a 180-mm-large steel-wire grid, and a grounded stainless-steel plate electrode. The radius of curvature of the needle tip was ~100 μm. The sharp needle and the steel-wire grid were connected to two Glassman FJ Series 120-watt high-voltage power supply units, respectively. The polymer under investigation was 125-μm-thick polyethylene terephthalate (PET) films (McMaster-Carr, Product Number 8567K52). In an ultrasonic cleaner, the PET films were rinsed in isopropyl alcohol (IPA) for 10 min and deionized (DI) water for another 10 min. Then, they were dried at 80 ºC for 1 day in a JeioTech ON-01E-120 oven. The films were sectioned into 152 mm or 76 mm large square samples.

During corona charging, a PET film sample was attached onto the grounded plate electrode. The steel-wire grid was in between the needle electrode and the film. The needle pointed to the film center. The distance between the needle tip and the grid was 40 mm. The distance between the grid and the polymer film was 4 mm. The voltage between the needle electrode and the grounded electrode ($V_{\text{Needle}}$) was -10 kV when the voltage on the grid ($V_{\text{Grid}}$) was -1 or -1.5 kV; $V_{\text{Needle}}$= -12 kV when $V_{\text{Grid}}$=-2, -2.5, or -3 kV. Corona was generated at 22 ºC for 1 min. After charging, the dipolar component of electric potential ($V_d$) was measured by a voltmeter (Trek Model-344 voltmeter), as the PET film rested on the grounded electrode plate (Fig.2a). A non-contact Kelvin probe was held 5 mm above the PET film, and scanned across the central area every 12.7 mm. The probe recorded the local average electric potential over a 10-mm-large area. Then, the sample was flipped and placed back to the plate electrode, and the $V_d$ measurement was repeated for the backside.

As the corona charged PET was removed from the grounded plate electrode, a small amount of free electrostatic charges would be produced on its surfaces, which greatly affected the surface voltage of the free-standing film ($V_f$). Without further treatment, $V_f$ could vary from -3 kV to 3 kV seemingly randomly. Flow charging was employed to control the surface density of the free charges ($\sigma_f$). Sodium formate (SF) (HCOONa) was provided by Sigma-Aldrich (Product No. 71539). It was dissolved in DI water. The molarity (*m*) was 0.3, 1, 3, 10, 30, or 100 mM. One side of a corona-charged 76×76 mm PET film was enclosed by a plastic box, with the edges sealed by duct tapes and Teflon tapes. The one-sidedly covered film was immersed in the SF solution, and the uncovered side was exposed to the liquid. The film was lifted out of the solution by a type-5582 Instron machine at a constant speed ($v_0$) of 4 mm/sec, as depicted in Figure 1(c). In addition to SF, we also tested solutions of sodium butyrate (SB) $CH_3(CH_2)_2COONa$ (Product No. B5887), sodium decanoate (SD) $CH_3(CH_2)_8COONa$ (Product No. C4151), and sodium stearate (SS) $CH_3(CH_2)_{16}COONa$ (Product No. S3381). All the salts were obtained from Sigma-Aldrich. They have the same cation (sodium) and the same carboxylate function group in the anion. The major difference is the anion chain length. The number of carbon-carbon bonds in anion varies from 0 to 17. In order to analyze the flow rate effect, for 10 mM SF solution, the lifting speed ($v_0$) was varied in the range from 0.01 to 10 mm/sec. The surface voltages of flow-electrified free-standing films were characterized on both the negative side ($V_{f,N}$) and the positive side ($V_{f,P}$) by the voltmeter along the centerline. The probe distance was 25 mm.

3. **Result and Discussions**

Figure 2(b) shows the typical dipolar component of electric potential of corona-charged film ($V_\text{d}$), measured from both the negative side and the positive side. The negative side was the surface exposed to the corona; the positive side was the back surface initially in contact with the grounded electrode plate. If the steel-wire grid was removed during corona charging, the voltage distribution fit well with the "bell jar" shape [25]. The magnitudes at the positive side and the negative side were similar. When the grid was in between the needle electrode and the film surface, it significantly homogenized the charge density, as illustrated in Figure 2(c). With $V_\text{Grid}$ being -2 kV, the positive side film voltage was ~2 kV, and the negative side voltage was ~ -2 kV. As shown in Figure 2(d), when $V_\text{Grid}$ varied from -1 kV to -3 kV, $V_\text{d}$ was changed nearly proportionally. Figure 2(e) shows the long-term stability of $V_\text{d}$ when $V_\text{Grid}$=-1.8 kV and $V_\text{Needle}$=-12 kV.

A finite element analysis (FEA) was conducted with COMSOL Multiphysics Electrostatic Module, to simulate the voltage distribution of a corona-charged film attached to a grounded plate electrode. The model geometry is shown in Figure 3(a). A 125-µm-thick 152-mm-large PET film was placed on a 100-mm-thick 200-mm-large steel electrode. The relative permittivity ($\varepsilon_r$) of PET and air were 3.0 and 1.0, respectively [26,27]. The charge densities on the positive and the negative sides were uniformly set to $\sigma_\text{d}$ and -$\sigma_\text{d}$, respectively. The voltage of the grounded electrode was 0. The voltage distribution is governed by the Poisson equation [28]

$$-\nabla(\varepsilon_0 \varepsilon_r \nabla V) = \rho \qquad (1)$$

where $V$ is the local voltage, $\varepsilon_0$ is the permittivity of vacuum, and $\rho$ is the spatial charge density. Free triangular mesh was employed. When $\sigma_\text{d}$ was 0.446 mC/m$^2$, as shown in Figure 3(b), the center point voltage 5 mm above the charged film was 2 kV at the positive side, and -2 kV at the negative side. In the central area more than 25 mm away from the edge, the boundary effect was secondary, and the voltage variation was less than 5%, consistent with the experimental data in Figure 2(c).

After the polymer film was lifted from the grounded electrode plate, to precisely control the free charges ($\sigma_\text{f}$), SF solution was utilized to further electrify one side of the corona-charged polymer. As shown in Figure 4(a) and (b), with increasing of the molarity of SF, $V_\text{f,N}$ and $V_\text{f,P}$ of the flow-electrified side was increased from -1.1 kV to 0; $V_\text{f,P}$ and $V_\text{f,N}$ of the covered side were increased from 0 to 1.1 kV and -2.3 kV to -1.1 kV, respectively, with the voltage difference between the positive side and the negative side ($\Delta V_f$) being nearly constant around 1.1~1.2 kV. To understand this phenomenon, the relationship between $\sigma_f$ and $V_f$ was analyzed based on the dipolar charge distribution model. The charge densities on the positive side and the negative side ($\sigma_P$ and $\sigma_N$) in Figure 4(a,b) were obtained from Eq. (1). The dipolar charges could be calculated as $\sigma_d = (\sigma_P - \sigma_N)/2$, and the free-charge density was $\sigma_f = \sigma_P + \sigma_N$. If $\sigma_P = -\sigma_N$, the electric potential was purely dipolar and $\sigma_f = 0$. As shown in Figure 4(c,d), $\sigma_d$ remained ~0.45 mC/m$^2$, consistent with the measurement result of $V_d$. The free-charge density ($\sigma_f$) was less than 0.1% of $\sigma_d$. It varied from -0.44 to 0.39 µC/m$^2$ in Figure 4(a) and -1.28 to -0.48 µC/m$^2$ in Figure 4(b), increasing with the SF molarity ($m$). As a result, $V_{f,N}$ and $V_{f,P}$ became higher on both sides. It is worth noting that $\Delta V_f$ and $\sigma_d$ were not affected by $m$ and $\sigma_f$.

FEA was performed to analyze the voltage distribution of a 125-µm-thick 76-mm-large free-standing PET film. The dipolar component of charge density was set to $\sigma_d$. Free charges ($\sigma_f$) were added either on the positive side or the negative side. The value of $\sigma_f$ and $\sigma_d$ in Figure 4(c,d) were used in the numerical calculation. Figure 4(e,f) show that as $\sigma_f$ was controlled by $m$, the valves of surface voltage largely varied. When $V_d = 2$ kV and $\sigma_f = 0.44$ µC/m$^2$, the voltages on the negative side of the film could be balanced to 0. This critical $\sigma_f$ will be referred to as the balanced free-charge density, $\sigma_{f,b}$.

The classical Gouy-Chapman model may explain the change of $\sigma_f$ upon flow electrification. Generally, when a PET film is exposed to an electrolyte solution, its surface would be negatively charged. The liquid-induced charge density is denoted by $\sigma_0$. It creates a positively charged electrical double layer [29], as shown in Figure 1(c). Within the double layer, there is a shear plane. If the liquid moves, the ions inside the shear plane would be adsorbed on the polymer surface [29–31]. When the flow rate is zero, for a non-charged film, $\sigma_0$ and the charges in the double layer are balanced. For a corona-charged film, the electrical double layer is influenced by both $\sigma_0$ and $\sigma_d$. Hence, when the flow rate is zero,

$$\sigma_{f,b} = \int_0^\infty \rho \, dz + \sigma_0. \tag{2}$$

where $z$ is the direction normal to the film surface. According to the Debye-Huckel approximation and the Poisson equation [28–30],

$$\rho = -\psi_0 \kappa^2 \varepsilon e^{-\kappa z}, \tag{3}$$

where $\psi_0$ represents the electric potential at the polymer surface, $\kappa^2 = [(e^2/\varepsilon k_B T)\sum_i 1000z_i^2 m_i N_A]$, $e$ represents the electron charge, $\varepsilon$ represents the liquid permittivity, $k_B$ represents the Boltzmann constant, $T$ represents temperature, $z_i$ represents the ion charge, $m_i$ represents the molarity of ion specie $i$, and $N_A$ is Avogadro's number. Combination of Eq. (2) and (3) gives

$$\sigma_0 - \sigma_{f,b} = \psi_0 \kappa \varepsilon \tag{4}$$

As the electrolyte solution flows across the solid surface, the free-charge density induced by the remnant ions can be calculated as

$$\sigma_f = \int_0^{z_s} \rho \cdot dz + \sigma_0 = \sigma_0^* e^{-z_s\sqrt{(e^2/\varepsilon k_B T)\sum_i 1000 z_i^2 m_i N_A}} + \sigma_{f,b} \tag{5}$$

where $\sigma_0^* = \sigma_0 - \sigma_{f,b}$. The trend of the testing data of $\sigma_f$ in Figure 4(c,d) can be captured by Eq. (5) quite well. For Fig.4(c), $\sigma_0^* = -1.1$ µC/m² and $z_s = 2.9$ nm; for Fig.4(d), $\sigma_0^* = -0.93$ µC/m² and $z_s = 3.0$ nm. When $z_s \to \infty$, $\sigma_f = \sigma_{f,b}$; when $z_s \to 0$, $\sigma_f = \sigma_0$. It can be seen that $\sigma_f$ tends to increase if $\sigma_{f,b}$, $m$, or the shear plane distance ($z_s$) becomes larger. The calculated $\sigma_f - m$ relationship is shown by the dashed lines in Figure 4(c,d).

In Figure 5(a,b), $V_{Grid}$ ranged from -1 kV to -3 kV, while the molarity of SF was kept at 10 mM and the lifting velocity ($v_0$) was 4 mm/s. The measurement data suggested that $V_f$ at the liquid-treated side was nearly constant at -0.6 kV. However, because $V_d$ depended on $V_{Grid}$, the electric potential at the untreated side varied in the range from 0.03 kV to 1.04 kV (Fig.5a) and from -2.16 to -1.04 kV (Fig.5b). The corresponding $\sigma_d$ increased linearly with $V_{Grid}$, as illustrated in Figure 5(c) and (d). Moreover, upon flow electrification, when $\sigma_d$ was increased, $\sigma_f$ increased on the negative side but decreased on the positive side, in agreement with Eq.(5).

In addition to the salt molarity, to accurately control $\sigma_f$, we also investigated the effects of the anion size and the flow velocity ($v_0$), as shown in Figure 6(a,b). The negative side of the corona-charged film was flow-electrified. The voltage difference between the two sides ($\Delta V_f$) was not affected by the liquid treatment. The magnitude of $V_{f,N}$ decreased as the anion size increased. Especially, 10 mM SS led to a near-zero $V_{f,N}$, suggesting that the voltage generated by the dipolar charges on the negative side was offset by the free charges. The factor of anion size came in by affecting $z_s$. A Ramé-hart Model-200 Contact Angle Goniometer was utilized to measure the contact angles of 100 mM SF, SB, SD, and SS solutions on a flat PET surface. The contact angles were 72°, 67°, 61°, and 55°, respectively. With a larger anion, the solid-liquid surface tension was greater [31]. That is, large anions were prone to being adsorbed and therefore, the shear plane was farther away from the solid surface. According to Eq. (5), with an increase in $z_s$, when $\sigma_0^*$ was negative, $\sigma_f$ and $V_f$ would be larger.

When the corona-charged film was pulled out of the salt solution faster, $V_{f,N}$ was decreased and converged to -0.7 kV when $v_0$=10 mm/s, as shown in Figure 6(b). When $v_0$ was 0.01 mm/s, the residual free-charge density ($\sigma_f$) was close to $\sigma_{f,b}$ and consequently, the overall surface voltage on the negative side ($V_{f,N}$) was nearly 0. Clearly, the charge separation across the shear plane was dependent on the relative motion. When $v_0 \rightarrow 0$, the shear plane was infinite far from the solid surface. As $v_0$ rose, the shear plane was closer to the film surface, so that $\sigma_f$ was decreased. Figure 6(c,d) indicates that under this condition, $\sigma_d$ would be unrelated to the anion size and the flow rate.

## 4. Conclusion

Corona-charged PET film possesses a stable dipolar component of electric potential ($V_d$). However, a relatively small surface density of free charges ($\sigma_f$) could greatly change the surface voltage of a free-standing film ($V_f$). We used flow electrification to control $\sigma_f$, as an aqueous solution of sodium salt flows across one side of a corona-charged PET film. The experimental results suggested that $V_f$ and $\sigma_f$ increased with the salt concentration ($m$) and the anion size, and the reduction in flow rate ($v_0$). This process did not influence $V_d$. When the grid voltage was higher, $V_f$ and $\sigma_f$ may either increased or decreased, depending on which side of the film was flow electrified. Equation (5) was derived from the classic Gouy-Chapman model, which explained the experimental observations quite well. The anion size and the flow rate affected $\sigma_f$ by changing the distance of the shear plane to the solid surface. These findings not only shed light on the details of ion behavior in diffuse layer at an electrified polymer surface, but also are critical to a wide variety of engineering applications, such as polymer electrets processing and treatment, bio-sensors and bio-actuators, electronics, energy harvesting, etc.

**Acknowledgement**

This study was supported by ARPA-E under Grant No. DE-AR0000737.

**Reference**

[1]  F. Galembeck, T. A. L. Burgo, Excess Charge in Solids: Electrets, in: Chem. Electrost., Springer International Publishing, Cham, 2017: pp. 91–106. https://doi.org/10.1007/978-3-319-52374-3_7.

[2]  S.B. Lang, D.K. Das-Gupta, A technique for determining the polarization distribution in thin polymer electrets using periodic heating, Ferroelectrics. 39 (1981) 1249–1252.

[3]  D.P. Erhard, D. Lovera, C. von Salis-Soglio, R. Giesa, V. Altstädt, H.-W. Schmidt, Recent advances in the improvement of polymer electret films, in: Complex Macromol. Syst. II, Springer, 2010: pp. 155–207.

[4]  R.M. Schaffert, C.D. Oughton, Xerography: A new principle of photography and graphic reproduction, JOSA. 38 (1948) 991–998.

[5]  A.S. Urban, A.A. Lutich, F.D. Stefani, J. Feldmann, Laser printing single gold nanoparticles, Nano Lett. 10 (2010) 4794–4798.

[6]  M. Paajanen, J. Lekkala, K. Kirjavainen, ElectroMechanical Film (EMFi) — a new

multipurpose electret material, Sensors Actuators A Phys. 84 (2000) 95–102. https://doi.org/10.1016/S0924-4247(99)00269-1.

[7]   S. Boisseau, G. Despesse, A. Sylvestre, Optimization of an electret-based energy harvester, Smart Mater. Struct. 19 (2010) 075015. https://doi.org/10.1088/0964-1726/19/7/075015.

[8]   Y. Zhong, R. Kou, M. Wang, Y. Qiao, Synthesis of centimeter-scale monolithic SiC nanofoams and pore size effect on mechanical properties, J. Eur. Ceram. Soc. 39 (2019) 2566–2573.

[9]   R. Kou, Y. Zhong, J. Kim, Q. Wang, M. Wang, R. Chen, Y. Qiao, Elevating low-emissivity film for lower thermal transmittance, Energy Build. 193 (2019) 69–77.

[10]  H. Wang, R. Kou, T. Harrington, K.S. Vecchio, Electromigration Effect in Fe-Al Diffusion Couples with Field-Assisted Sintering, Acta Mater. (2020).

[11]  A. Oliveros, A. Guiseppi-Elie, S.E. Saddow, Silicon carbide: a versatile material for biosensor applications, Biomed. Microdevices. 15 (2013) 353–368. https://doi.org/10.1007/s10544-013-9742-3.

[12]  M.C. Wiles, T. VanderNoot, D.J. Schiffrin, Electrosorption of amphiphilic affinity dyes at liquid/liquid interfaces for biosensor applications, J. Electroanal. Chem. Interfacial Electrochem. 281 (1990) 231–244.

[13]  S. Wang, J. Xu, W. Wang, G.J.N. Wang, R. Rastak, F. Molina-Lopez, J.W. Chung, S. Niu, V.R. Feig, J. Lopez, T. Lei, S.K. Kwon, Y. Kim, A.M. Foudeh, A. Ehrlich, A. Gasperini,

Y. Yun, B. Murmann, J.B.H. Tok, Z. Bao, Skin electronics from scalable fabrication of an intrinsically stretchable transistor array, Nature. 555 (2018) 83–88. https://doi.org/10.1038/nature25494.

[14] B. Tabti, L. Dascalescu, M. Plopeanu, A. Antoniu, M. Mekideche, Factors that influence the corona charging of fibrous dielectric materials, J. Electrostat. 67 (2009) 193–197. https://doi.org/10.1016/J.ELSTAT.2009.01.047.

[15] P. Molinié, Charge injection in corona-charged polymeric films: potential decay and current measurements, J. Electrostat. 45 (1999) 265–273. https://doi.org/10.1016/S0304-3886(98)00053-9.

[16] G. Chen, A new model for surface potential decay of corona-charged polymers, J. Phys. D. Appl. Phys. 43 (2010) 055405. https://doi.org/10.1088/0022-3727/43/5/055405.

[17] Y. Zhong, R. Kou, M. Wang, Y. Qiao, Electrification Mechanism of Corona Charged Organic Electrets, J. Phys. D. Appl. Phys. (2019).

[18] V.N. Kestelman, L.S. Pinchuk, V.A. Goldade, Electrets in engineering: fundamentals and applications, Springer Science & Business Media, 2013.

[19] A. Jaworek, A. Krupa, T. Czech, Modern electrostatic devices and methods for exhaust gas cleaning: A brief review, J. Electrostat. 65 (2007) 133–155.

[20] H.F. Richards, The contact electricity of solid dielectrics, Phys. Rev. 22 (1923) 122.

[21] W.D. Rees, Static hazards during the top loading of road tankers with highly insulating

liquids: flow rate limitation proposals to minimize risk, J. Electrostat. 11 (1981) 13–25.

[22] K. Manna, C.-Y. Hsieh, S.-C. Lo, Y.-S. Li, H.-N. Huang, W.-H. Chiang, Graphene and graphene-analogue nanosheets produced by efficient water-assisted liquid exfoliation of layered materials, Carbon N. Y. 105 (2016) 551–555.

[23] G. Touchard, Flow electrification of liquids, J. Electrostat. 51 (2001) 440–447.

[24] R. Kou, Y. Zhong, Y. Qiao, Effects of anion size on flow electrification of polycarbonate and polyethylene terephthalate, Appl. Phys. Lett. 115 (2019) 73704.

[25] T.-J. Wang, Y. Wei, Y. Liu, N. Chen, Y. Liu, J. Ju, H. Sun, C. Wang, H. Lu, J. Liu, Direct observation of laser guided corona discharges, Sci. Rep. 5 (2015) 18681.

[26] G.G. Raju, Dielectrics in electric fields, Marcel Dekker, 2003.

[27] J. Artbauer, Electric strength of polymers, J. Phys. D. Appl. Phys. 29 (1996) 446.

[28] N. Jonassen, Electrostatics, Springer Science & Business Media, 2013.

[29] S.S. Dukhin, B.V. Deriaguine, Surface and Colloid Science: Electrokinetic Phenomena: Translated from the Russian by A. Mistetsky and M. Zimmerman, Plenum Press, 1974.

[30] P.C. Hiemenz, R. Rajagopalan, Principles of Colloid and Surface Chemistry, revised and expanded, CRC press, 1997.

[31] R.J. Hunter, Foundations of colloid science, Oxford university press, 2001.

**Figures**

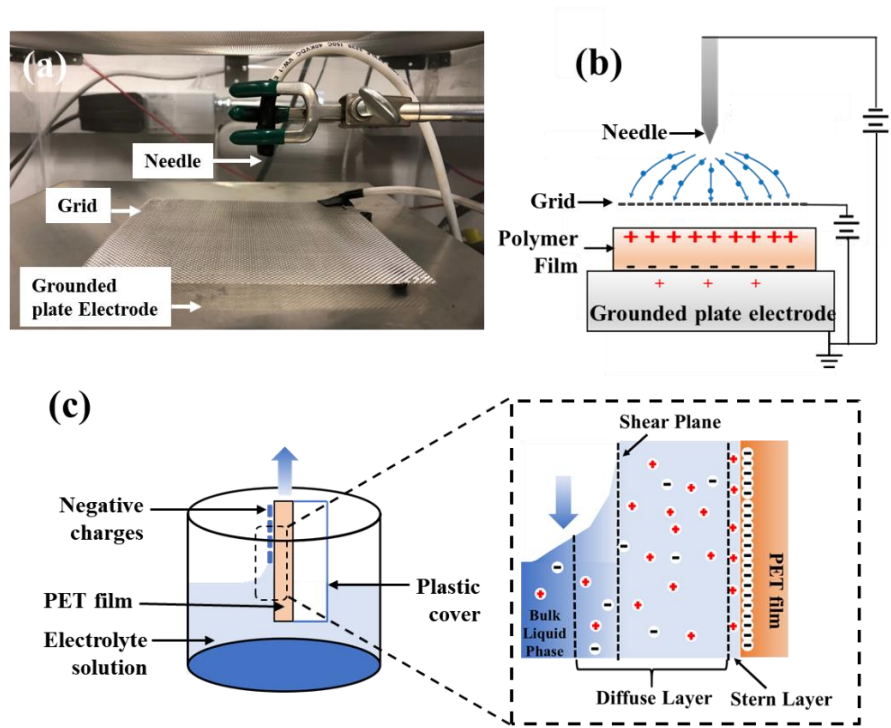

**Figure 1** a) Photo and b) schematic of the corona charging system. c) Schematic of the flow electrification setup. The magnified view on the right shows that a liquid flow breaks the electric equilibrium at the PET surface.

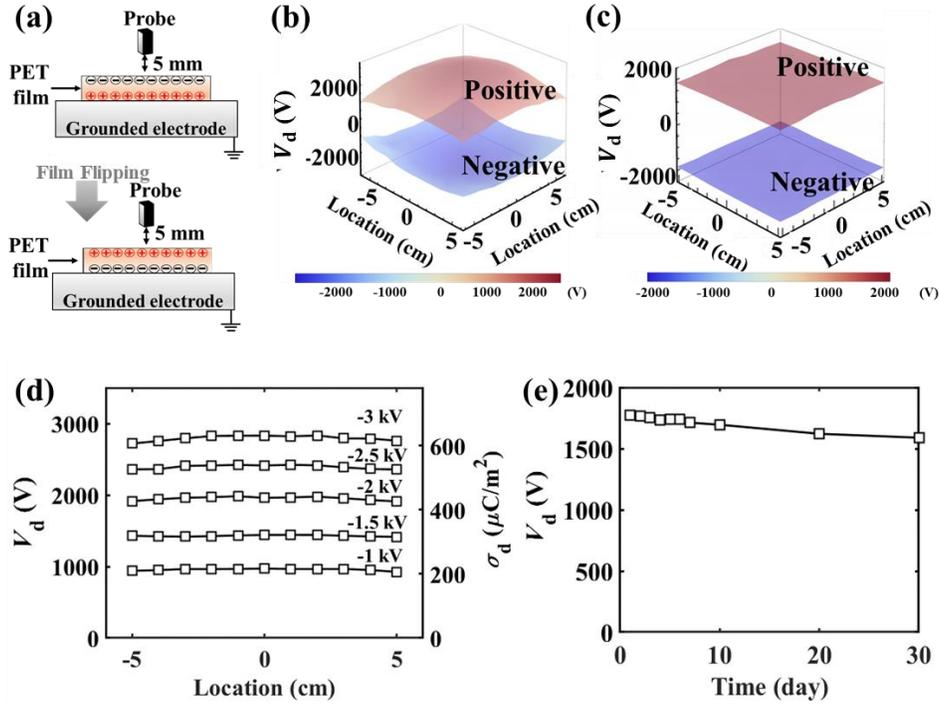

**Figure 2** a) Schematic of the measurement of the dipolar component of electric potential ($V_d$). b,c) respectively show the $V_d$ distributions of a 125-μm-thick PET film corona-charged without and with the steel-wire grid ($V_{Grid}$= -2 kV and $V_{Needle}$=-12 kV). "Positive" and "Negative" indicate the measurement direction. d) The measured $V_d$ and the associated surface charge density ($\sigma_d$) of the PET film; the numbers next to the curves indicate $V_{Grid}$ in kV. e) The stability of $V_d$ of a corona-charged PET film ($V_{Needle}$ = -12 kV and $V_{Grid}$=-1.8 kV). $V_d$ is measured along the centerline with the probe distance of 5 mm. The PET film size is 152 mm by 152 mm.

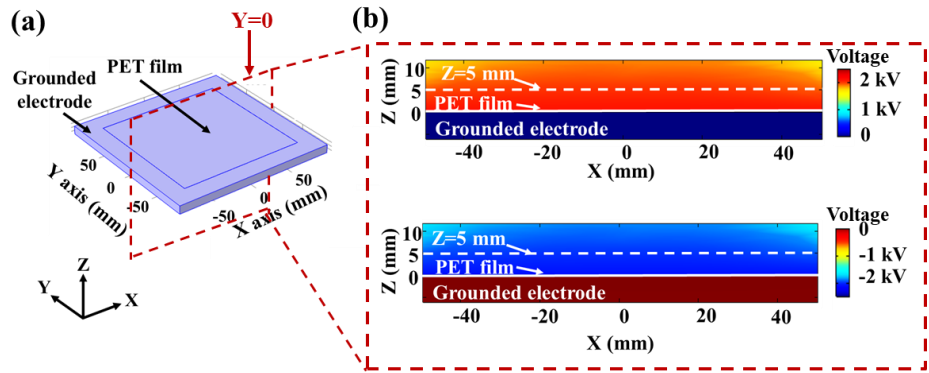

**Figure 3** a) The finite element analysis (FEA) model. The center point of the top surface of the PET film is at (0,0,0). The bottom surface is attached onto a grounded electrode plate. One side of the PET film is charged negatively with the surface charge density of $-\sigma_d$; the other side is positively charged, with the same amplitude of surface chare density ($\sigma_d$). b) Calculated voltage distributions at plane Y = 0 on the positive side (top) and the negative side (bottom). $\sigma_d$ is 0.446 mC/m$^2$. Electric potentials at Z= 5 mm are nearly 2 kV (top) and -2 kV (bottom) at the positive side and the negative side, respectively.

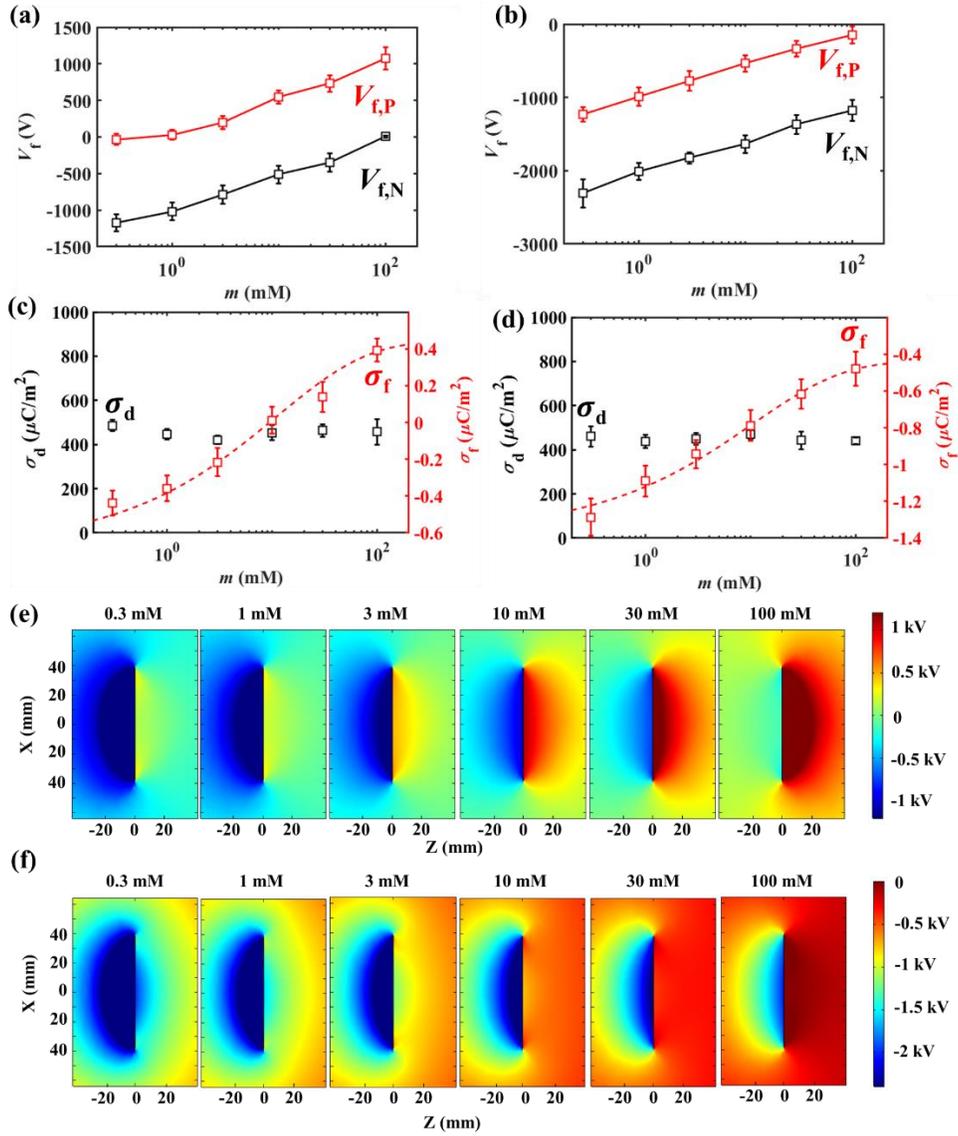

**Figure 4** The effects of the SF concentration ($m$) on $V_f$, with the $V_{Grid}$ and $V_{Needle}$ being -2 kV and -12 kV, respectively: a) the negative side is flow electrified; b) the positive side is flow electrified. c,d) show $\sigma_f$ and $\sigma_d$ for a,b), respectively. e,f) Calculated voltage distributions at plane Y = 0 of a PET film, as the liquid electrifies its negative side and the positive side, respectively. The polymer surface is vertical to the Z axis and the center point of its surface is located at (0,0,0). The liquid is a SF solution, with the SF molarity varying from 0.3 mM to 100 mM, as shown at the top of each chart. The PET film size is 76 mm by 76 mm.

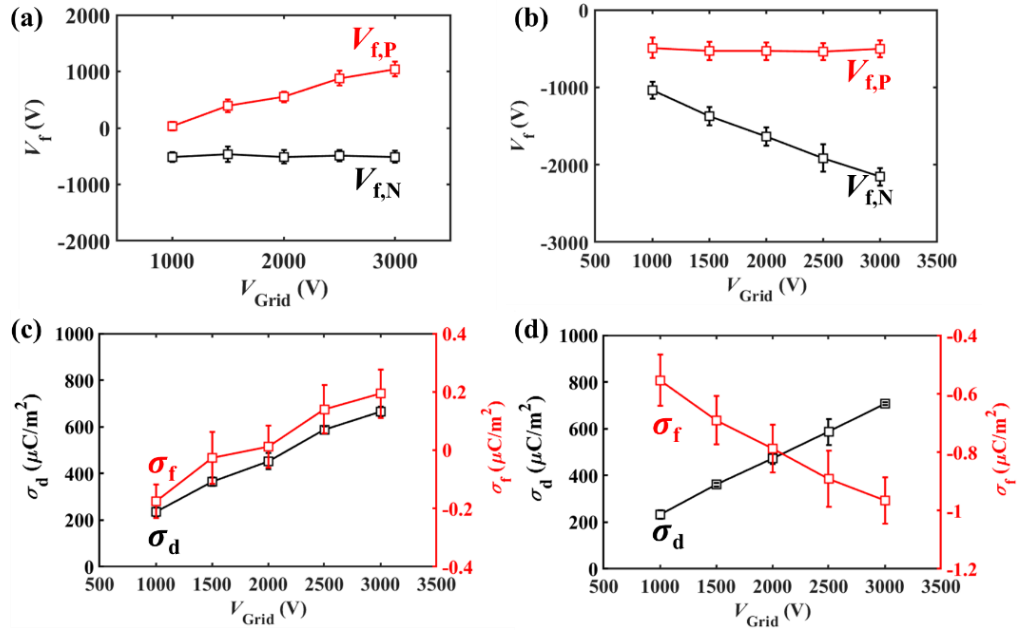

**Figure 5** $V_f$ of PET films flow-electrified by a 10 mM SF solution. The films have been corona-charged with various grid voltages ($V_{Grid}$): a) The negative side is flow electrified; b) the positive side is flow electrified. c,d) show the corresponding $\sigma_f$ and $\sigma_d$ of a,b), respectively. $V_{Needle}$ = -10 kV when $V_{Grid}$=-1 or -1.5 kV; $V_{Needle}$= -12 kV when $V_{Grid}$=-2, -2.5, or -3 kV; $v_0 = 4$ mm/sec.

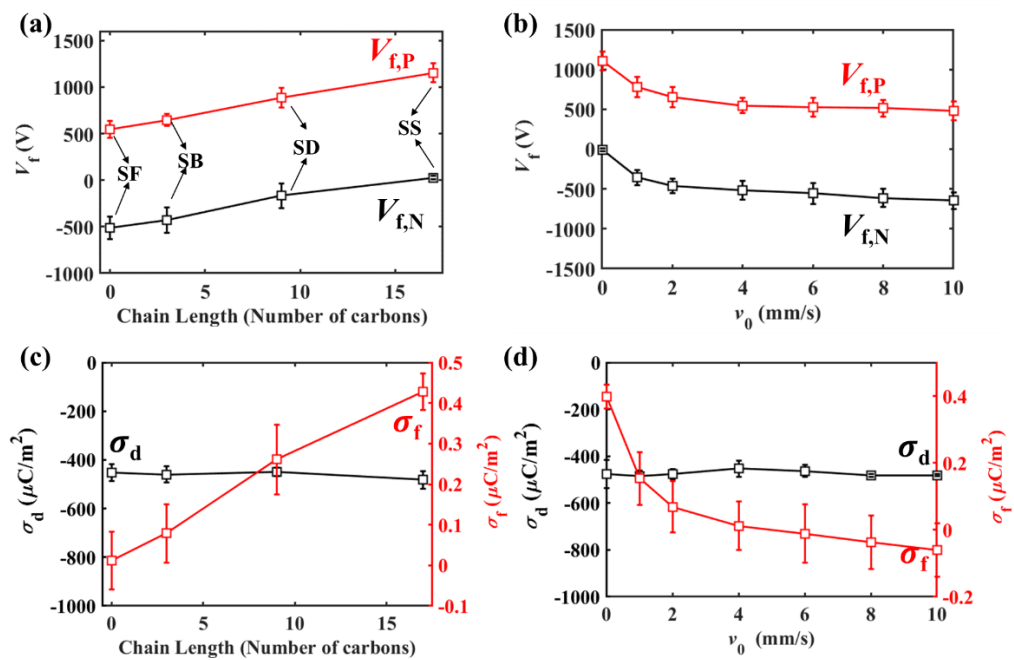

**Figure 6** (a) $V_f$ of corona-charged PET films treated by 10 mM SF, SB, SD, and SS solutions. (b) The flow rate ($v_0$) influences $V_f$ of SF-treated PET films; the negative side is flow-electrified. The difference between $V_{f,P}$ and $V_{f,N}$ remains nearly constant 1.1~1.2 kV. (c,d) show the corresponding $\sigma_d$ and $\sigma_f$ of (a,b), respectively.